\documentclass[10pt,twocolumn,letterpaper]{article}

\usepackage{cvpr}
\usepackage{multirow}
\usepackage{times}
\usepackage{xcolor}
\usepackage{epsfig}
\usepackage{subcaption}
\usepackage{graphicx}
\usepackage{amsmath}
\usepackage{amssymb}
\usepackage{booktabs}

\usepackage{float}

\cvprfinalcopy 

\usepackage{hyperref}
\usepackage[ruled]{algorithm2e}
\newcommand{\KL}{\textrm{KL}}
\newcommand{\ELBO}{\textrm{ELBO}}
\newcommand{\diag}{\textrm{diag}}

\ifcvprfinal\fi
\begin{document}
\title{Unsupervised Lesion Detection via Image Restoration \\with a Normative Prior}

\author{Xiaoran Chen$^1$\thanks{Corresponding author. This work was partly supported by the Swiss National Science Foundation under grant 205321\_173016 and Platform for Advanced Scientific Computing under grant HPC-Predict.}\hspace{10mm}Suhang You$^2$\thanks{Work done while at ETH Z\"urich.}\hspace{10mm}Kerem Can Tezcan $^1$\hspace{10mm}Ender Konukoglu$^{1}$\\[2mm]
$^1$Computer Vision Laboratory, ETH Z\"urich\hspace{10mm} 
$^2$ARTORG, University of Bern\\[-1.5pt]
\tt\small$^1$ \{chenx,tezcan,ender.konukoglu\}@vision.ee.ethz.ch
\tt\small$^2$suhang.you@artorg.unibe.ch }

\maketitle

\begin{abstract}
Unsupervised lesion detection is a challenging problem that requires accurately estimating normative distributions of healthy anatomy and detecting lesions as outliers without training examples. 
Recently, this problem has received increased attention from the research community following the advances in unsupervised learning with deep learning. Such advances allow the estimation of high-dimensional distributions, such as normative distributions, with higher accuracy than previous methods.
The main approach of the recently proposed methods is to learn a latent-variable model parameterized with networks to approximate the normative distribution using example images showing healthy anatomy, perform \emph{prior-projection}, i.e. reconstruct the image with lesions using the latent-variable model, and determine lesions based on the differences between the reconstructed and original images. 
While being promising, the prior-projection step often leads to a large number of false positives.
In this work, we approach unsupervised lesion detection as an image restoration problem and propose a probabilistic model that uses a network-based prior as the normative distribution and detect lesions pixel-wise using MAP estimation. 
The probabilistic model punishes large deviations between restored and original images, reducing false positives in pixel-wise detections. 
Experiments with gliomas and stroke lesions in brain MRI using publicly available datasets show that the proposed approach outperforms the state-of-the-art unsupervised methods by a substantial margin, +0.13 (AUC), for both glioma and stroke detection. Extensive model analysis confirms the effectiveness of MAP-based image restoration.

\end{abstract}



\section{Introduction}
\label{sec1}
Detecting lesions plays a critical role in radiological assessment; it is often the first step in the diagnosis pipeline. 
Manual detection, the dominant current practice, relies on an exceptional understanding of the normal anatomy. 
Specifically, radiologists detect lesions as regions that deviate from the normal variation of the healthy anatomy, and then identify the disease based on the features of the lesion as well as other patient information. This detection process is unsupervised in the sense that the radiologist is not looking for a specific lesion but for any abnormal variation.

Developing algorithmic approaches for automatic unsupervised lesion detection is important for both algorithm development and clinical practice. For algorithm development, accurate unsupervised lesion detection would allow developing further algorithms that are robust to lesions unseen in the training set. For instance, a machine learning-based segmentation algorithm that is aware of the presence of an outlier can take this into account while segmenting the rest of the image avoiding being affected by the abnormal feature responses of the outlier area. Such robustness would be extremely helpful for a variety of tasks researchers are tackling with machine learning methods, such as segmentation, localization, reconstruction and restoration. Good unsupervised lesion detection methods can also be applied to detecting artifacts and furthermore developing methods robust to such artifacts. On the other hand, for clinical practice, a generic and accurate unsupervised lesion detection would be a useful pre-screening tool to improve efficiency in radiological assessment by focusing expert effort directly on abnormal areas. This aspect is becoming particularly critical considering the increase in image resolution and number of modalities used in clinical routines, which on one hand have the potential to improve diagnostic accuracy but on the other hand result in drastically larger image volumes and number of images to be examined for each study, thus more effort. 

Machine learning-based approaches have attracted considerable attention for lesion detection in the last decade. Majority of the research effort in this direction focus on developing supervised algorithms~\cite{ayachi2009brain,zikic2012context,geremia2011spatial,dong2017automatic,pereira2016brain,kamnitsas2017efficient,li2018fully}. 
In this approach, algorithms are optimized during the training phase to detect lesions of \emph{pre-specified} types based on examples in the training set. 
When tested on the same lesion types, these methods yielded state-of-the-art detection and segmentation performance. 
However, being optimized to detect pre-specified lesion types limits their applicability to unseen lesions.
In theory, adding examples of all possible types of lesions in training can address this issue but this option, even when feasible, would be very expensive in manual effort and time.

Unsupervised lesion detection methods take another approach. They focus on learning prior knowledge on the appearance of healthy anatomy in order to detect lesions as areas that disagree with the prior knowledge, mimicking radiologists.
In one of the first works, a probabilistic atlas of healthy anatomy is utilized as the prior knowledge and built tissue-specific mixture models with an outlier class to segment healthy tissue while identifying areas that do not fit the other tissue classes~\cite{van2001automated}.
In similar approaches,  an atlas with spatial features instead of intensities is used~\cite{moon2002automatic} and combines spatial and intensity atlases~\cite{prastawa2004brain}. 
More recent works use prior models of patches of images to include contextual information around a pixel in the detection process. 
Considering the curse of dimensionality in modeling high dimensional distributions, such works adopt dimensionality reduction methods, such as principle component analysis~\cite{zacharaki2012abnormality,erus2014individualized}, patch-based mixture model~\cite{cardoso2015template} and sparse representation~\cite{zeng2016abnormality}. 

Advances in deep learning (DL)-based unsupervised learning methods provided strong alternatives for approximating distributions in high dimensions. 
Neural samplers, such as Generative Adversarial Networks (GAN)~\cite{goodfellow2014generative}, can learn to generate realistic high-dimensional images similar to the ones seen in the training set starting from a lower-dimensional latent space, and DL-based non-linear latent variable models, such as Variational Autoencoders (VAE)~\cite{kingma2013auto}, approximate the distribution from the samples in a training set, using networks to map from images to latent space and back, implementing the original idea~\cite{mackay1995bayesian} efficiently. 
DL-based approaches yield better approximations to the high-dimensional distributions, as evidenced by the realistic samples they can generate. 
Recently proposed works~\cite{schlegl2017unsupervised, baur2018deep, Pawlowski2018UnsupervisedLD, chen2018unsupervised} already apply DL-based models to approximating normative distributions with only images showing healthy anatomy and then detect lesions as the outliers. 
These methods typically has three steps. First, a given image with a lesion is projected to the latent space of the model, which in other words is to find the latent space representation of the image. 
Since the models have only seen healthy anatomy during training, the most likely latent representations correspond to images with healthy anatomy.
Based on this assumption, the original image is reconstructed from its latent representation, with the lesion reconstructed with large errors and the rest of the image reconstructed faithfully. 
Hence, the difference between the reconstruction and the original image will highlight the lesion. 
While using this \emph{prior projection} is a sound idea, it relies on the assumption that healthy areas in the images will remain the same in the reconstructed image, meaning that the latent space representations of the image with lesion and its lesion-free version are very similar. Unfortunately, this assumption may not hold since the intensities in the lesion area may greatly affect the projection step, causing large deviations between the mentioned latent space representations. In contrast, a method that directly takes such possible deviations into account can lead to improved detection accuracy.

In this work, we propose an approach that casts the unsupervised detection problem as an image restoration problem. 
The proposed method also uses DL-based non-linear latent variable models to approximate the normative distribution of healthy anatomy from example images.
However, instead of using prior projection, we formulate the detection as a Maximum-A-Posteriori (MAP) image restoration problem with the normative image prior estimated by a DL-based model, in the same spirit as the method~\cite{tezcan2018mr}, and a likelihood model that places minimal assumptions on the lesion. 
The MAP estimation is then solved using gradient ascent. 
The proposed image restoration method is agnostic to the choice of the priors and therefore can be applied with any DL-based latent variable model. 
To learn the image prior for image restoration, we use VAE as well as GMVAE~\cite{dilokthanakul2016deep,johnson2016structured}, an extension of VAE that uses Gaussian mixture modeling in the latent space. 
We evaluate and analyze the proposed method on two different Magnetic Resonance Imaging (MRI) lesion datasets, one showing brain tumors and the other showing lesions due to stroke.
A preliminary version of this work~\cite{you2019unsupervised} was presented at International Conference on Medical Imaging with Deep Learning. In this extended version, we present an in-depth analysis of the model and evaluations on an additional dataset. 

We note that unsupervised lesion detection approaches, including the method proposed here, currently yield lower detection accuracy compared to the supervised approaches. This is due to the fact that unsupervised lesion detection is a much more challenging task. The method we present here improves the state-of-the-art in unsupervised lesion detection and closes the accuracy gap between supervised and unsupervised approaches a bit further, taking a step towards making unsupervised lesion detection a viable alternative to supervised approaches with important benefits as described above. 

\begin{figure}[ !httb]
    \begin{subfigure}[b]{0.45\textwidth}
    \includegraphics[scale=0.38]{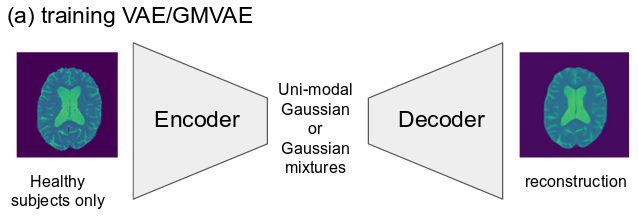}
    \label{fig:overview_training}
    \end{subfigure}
    
    \begin{subfigure}[b]{0.45\textwidth}
    \includegraphics[scale=0.38]{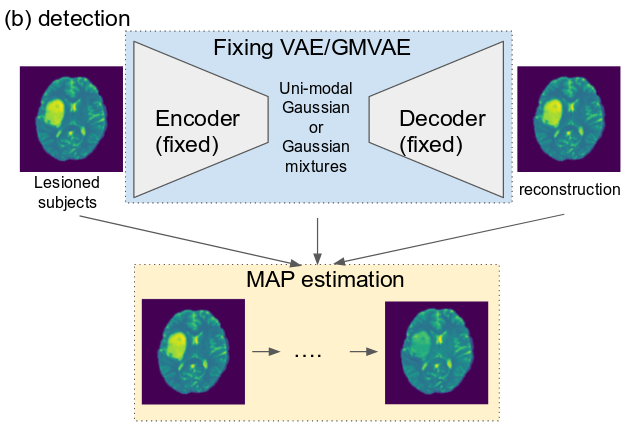}
    \label{fig:overview_detection}
    \end{subfigure}
    \caption{Overview of the detection method. (a) Training VAE/GMVAE as the image prior model. (b) Restoring lesioned subjects with the image prior model and MAP estimation. The images shown above are for the illustration purpose only.}
    \label{fig:overview}
\end{figure}

\section{Method}
\label{sec: Method}
We start this section with a brief review on normative distribution estimation using latent variable models, particularly VAE and GMVAE, and move on to give more details on MAP-based image restoration for unsupervised lesion detection. For an overview of the method, please refer to Figure \ref{fig:overview}

\subsection{Learning Prior Normative Distribution}

Latent variables models have been a popular choice for approximating high-dimensional distributions from a set of examples. 
Let us denote an image with $N$ pixels with $X\in\mathbb{R}^N$, a latent variable model for the distribution of $X$ in its most generic form is given as
\begin{equation*}
    P(X) = \int P(X|z) P(z) dz,
\end{equation*}
where $z\in\mathbb{R}^M$ is the latent variable, $P(z)$ is the prior distribution in the latent space. $M$ is the dimension of the latent space and often a much smaller $M$ than $N$ is used, i.e. $M\ll N$, assuming images can be represented in a lower dimensional sub-space of $\mathbb{R}^N$. The conditional distribution $P(X|z)$ encodes the mapping between the latent space and the image space.

The simplest example of a latent variable model is the probabilistic principal component analysis proposed~\cite{tipping1999probabilistic}, where $P(X|z)$ encodes a linear map between the latent and image spaces within a Gaussian distribution, i.e. $P(X|z) = \mathcal{N}\left(X; z^TU, \sigma^2\right)$ with $U\in \mathbb{R}^{M\times N}$ and $\sigma$ variance of the assumed noise in the data. Being powerful already, this model has been used as a normative distribution in unsupervised lesion detection~\cite{erus2014individualized}. However, a linear mapping can be limiting and a non-linear latent variable model with the mapping parameterized by neural networks~\cite{mackay1995bayesian} has been proposed to address the limitations. The version of this model using Gaussian conditional distributions can be given as
\begin{equation*}
    P(X|z) = \mathcal{N}\left(X; \mu_X(z; \theta), \Sigma_X(z; \theta)\right),
\end{equation*}
where $\mu_X(z; \theta)$ and $\Sigma_X(z;\theta)$ are neural networks and $\theta$ represents the entire set of parameters. Given this model, learning is modeled as maximizing the evidence $\max_{\theta} \sum_n \log P(X_n)$, where the subscript $n$ goes over the samples. 

The optimization of $P(X)$ requires solving the integral over $z$ either analytically or numerically. The former is not possible and the latter becomes infeasible for even modestly large $M$. MacKay in his article proposes to use importance sampling to address the issue~\cite{mackay1995bayesian}. VAE~\cite{kingma2013auto} takes this approach and defines a distribution $Q(z|X)$ parameterized with another neural network to approximate the true posterior $P(z|X)$, and maximizes the evidence lower bound (ELBO) rather than the evidence itself
\begin{eqnarray*}
    \log P(X) &\geq& \mathbb{E}_{Q(z|X)}\left[\log P(X|z)\right] - \KL\left[Q(z|X)\| P(z)\right]\\
    \ELBO(X) &=& \mathbb{E}_{Q(z|X)}\left[\log P(X|z)\right] - \KL\left[Q(z|X)\| P(z)\right],
\end{eqnarray*}
where $P(X|z)$ and $Q(z|X)$ are modeled with two networks, with the former parameterized with $\theta$ and the latter with $\phi$, and $\KL$ represents the Kullback-Leibler divergence. In VAE, learning is defined as 
\begin{equation*}
    \max_{\theta, \phi} \sum_n \ELBO(X_n),
\end{equation*}
and is implemented with stochastic sampling to compute the expectations in the ELBO. For further information regarding the optimization of the VAE objective function, we refer the readers to the original paper~\cite{kingma2013auto}. 

Following the modeling choices of the original work, here we use the VAE model with the following
\begin{enumerate}
    \item[-] $P(X|z)=\mathcal{N}\left(X; \mu_X(z), \diag\left(\sigma^2_X\right) \right)$, 
    \item[-] $Q(z|X)=\mathcal{N}\left(z; \mu_z(X), \diag\left(\sigma^2_z(X)\right) \right)$, and 
    \item[-] $P(z) = \mathcal{N}\left(0, \mathbf{I}_M\right)$,
\end{enumerate}
where $\diag(\sigma^2_X)$ denotes a diagonal matrix with $\sigma^2_X$ elements on the diagonal. $\mathbf{I}_M$ is the identity matrix in $M$ dimensions, and $\mu_X(z), \mu_z(X)$ and $\sigma_z(X)$ are parameterized as neural networks, whose architectures are described in Section~\ref{sec:experiments}. 

\noindent \emph{Remark:} Note that with the Gaussian distribution models, $\log P(X|z)$ is a computed as $\ell_2$ loss between the image $X$ and its mean reconstruction. The $\ell_2$ loss is derived from Gaussian distribution with the mean $\mu_X(z)$ and the homoscedatic variance $1 / \sigma^2_X$. The variance is set as a constant value of $\sqrt{\frac{1}{2}}$ for each pixel. The approaches that uses prior projection with VAE, or other latent-variable models, often first project the image to its latent representation $z$ and and then reconstruct the image with $\mu_X(z)$. The difference between $\mu_X(z)$ and $X$ is then assumed to be the lesion.

\begin{figure*}
    \centering
    \includegraphics[scale=0.4]{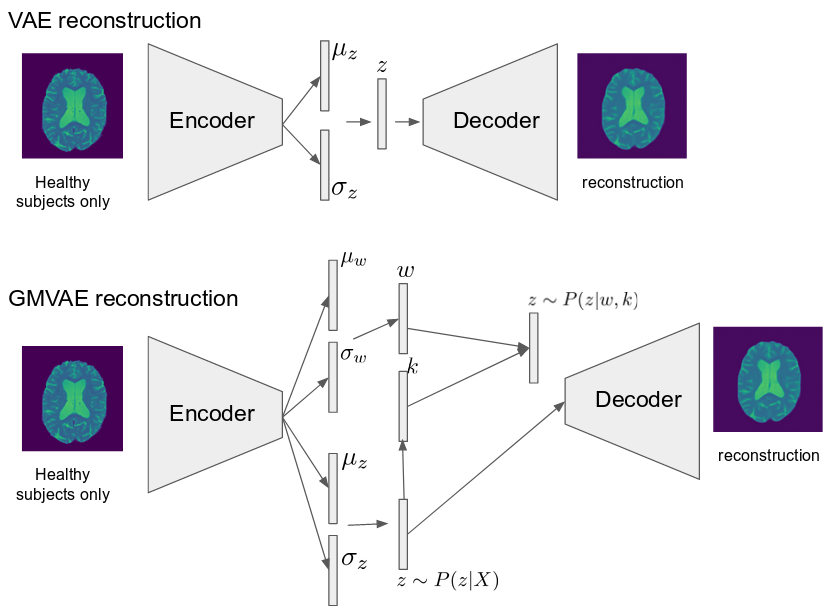}
    \caption{Overview of the reconstruction process of VAE/GMVAE. Images are only for illustration purpose.}
    \label{fig:reconstruction}
\end{figure*}

The original VAE model uses a unimodal distribution as the prior in the latent space. A more expressive prior can potentially allow the latent variable model to fit more complicated distributions to the samples. With this motivation,  Gaussian Mixture VAE (GMVAE)~\cite{dilokthanakul2016deep} is proposed and uses Gaussian mixture models as the prior in the latent space.
 
To model the latent distribution as Gaussian mixtures, the authors introduce two additional latent variables in GMVAE: $k$ for mixture assignment and $\omega$ for mixture model coefficients. The prior distribution in the latent space is given as

\begin{equation}
 P(z|\omega,k)=\prod_{c=1}^C \mathcal{N}\left(\mu_{k_c}(\omega), \diag(\sigma^2_{k_c}(\omega))\right)^{k_c},
\end{equation}
where $c$ is the pre-specified number of Gaussian mixture components, $k\sim \text{Mult}(\frac{1}{c})$ is a one-hot vector and $\omega \sim N(0,\mathbf{I}_M)$.
$\mu_{c_k}(\omega)$ and $\sigma^2_{k_c}(\omega)$ are functions of $\omega$ parameterized as neural networks. 

Similar to VAE, an ELBO can be derived for GMVAE~\cite{dilokthanakul2016deep}, using multiple approximations for the posterior distributions
\begin{eqnarray}
     \label{eq:gmelbo}\ELBO(X) &=& \mathbb{E}_{Q(z|X)}\left[\log P(X|z)\right]\\
     \nonumber &-& \mathbb{E}_{Q(\omega|X)P(k|z,\omega)} \left[\KL\left[Q(z|X)|| P(z|\omega,k)\right]\right]\\
     \nonumber &-& \KL\left[Q(\omega|X)||P(\omega)\right]\\
     \nonumber &-& \mathbb{E}_{Q(z|X)Q(\omega|X)}\left[\KL\left[P(k|z,\omega)||P(k)\right]\right].
\end{eqnarray}
The mixture model in the latent space gives rise to two additional distributions $Q(\omega | X)$ and $P(c|z, \omega)$. We use a Gaussian distribution with diagonal covariance for the first one, $Q(\omega|X)=\mathcal{N}\left(\omega; \mu_{\omega}(X), \diag\left(\sigma^2_{\omega}(X)\right)\right)$, while the posterior for $c$ can be computed analytically for a given $\omega$ and $z$. We use the same models for $P(X|z)$ and $Q(z|X)$ as in the VAE model. Training of the GMVAE model follows a similar approach as VAE, and we refer the interested reader to the original paper~\cite{dilokthanakul2016deep} for details.

We use convolutional neural networks to parameterize the necessary distributions in the VAE and the GMVAE models and train the models with MR images acquired from healthy individuals to obtain the normative distribution, which is used in the restoration framework as explained next. The details on the architectures and datasets are presented in Section~\ref{sec:experiments}. While the VAE and GMVAE models are used for demonstrations, the MAP-based outlier detection described next can be combined with any latent variable models, such as the flow-based unsupervised learning approaches~\cite{grathwohl2018ffjord}. An illustrated reconstruction process can be found in Figure \ref{fig:reconstruction}.

\subsection{Detecting outliers with MAP-based restoration}

The restoration framework assumes the image with the lesion is a ``normal'' image, i.e. one that is coming from the normative distribution, but with an additional ``noise'' component, the lesion. The goal is to restore the normal image from the ``noisy'' observation and in the meanwhile detect the lesion as noise. This is fundamentally different from the prior projection model or detecting outliers in the latent space with another metric. The approach here specifically aims to retain the normal anatomy in the images and only change the outlier lesion part during the restoration. Next we describe how we achieve this with a probabilistic model and MAP estimation. 

Let us denote an image with a lesion as $Y \in \mathbb{R}^{N}$. 
We assume that $Y$ is a noisy version of $X\in\mathbb{R}^N$, $Y = X + D$, modeling the lesion with $D\in\mathbb{R}^N$ and the noise-free, which corresponds to lesion-free, version of $Y$ with $X$.
In areas of the image where no lesion is present, $D=0$ should hold at those pixels. 
In cases with no lesion at all,  $Y=X$ should hold in the entire image, i.e. the model should find no lesion. 

The usual MAP estimation maximizes the posterior distribution of $X$ given $Y$ 
\begin{equation*}
    \arg\max_X \log P(X|Y) = \arg\max_X \left[\log P(Y|X) + \log P(X)\right],
\end{equation*}
where $P(Y|X)$ is the likelihood term, which can be interpreted as the data consistency term. This term reflects the assumptions on $D$, which we detail in Section~\ref{DataConsistency}. $P(X)$ is the normative prior, i.e. distribution of healthy images. These two terms form a balanced optimization problem. The first term aims to preserve the image by penalizing deviations between the observed image and its restored version $X$. Maximizing that term alone would not change the image. The second term, however, yields higher values for images that fit the normative distribution. An image with lesions has a lower probability in the normative distribution and consequently gives a lower $\log P(X)$. For such images, the optimal point is to remove the outlier lesion that does not fit the normative distribution, assigning it to $D$, while keeping the normal anatomy fixed between $Y$ and $X$. For an image without lesions, leaving the image unchanged is the optimal solution. 

The MAP estimation optimizes $\log P(X)$, but this term is neither analytically tractable nor easy to compute via sampling for most non-linear latent variable models. Instead, for VAE and GMVAE, we propose to optimize the evidence lower bound, or in other words, to solve the approximated MAP estimation problem
\begin{equation}\label{eq:restormap2}
    \arg\max_X \log P(X|Y) \approx \arg\max_X \left[\log P(Y|X) + \ELBO(X)\right].
\end{equation} 
The difference between $\ELBO(X)$ and $\log P(X)$ is given with $\KL\left[Q(z|X) \| P(z|X)\right]$. Therefore, as the approximation $Q(z|X)$ gets better, the KL-divergence will go to zero and maximizing $\ELBO$ will get closer to maximizing $\log P(X)$. An important advantage of using $\ELBO(X)$ is that, it is formed of differentiable functions allowing for a gradient-based optimization scheme to solve the approximate MAP problem.

To optimize the objective formulated in Equation~\ref{eq:restormap2}, we adopt a gradient ascent method and perform iterative optimization to obtain the restored image. Specifically, in each iteration, we approximate the gradients of $P(X|Y)$ by taking the gradient of Equation \eqref{eq:restormap2} 
\begin{align}
    G_{X^i} &=  \frac{\partial \left[\log P(Y|X) + \ELBO(X)\right]}{\partial X} \big\rvert_{X=X^i},\\
    X^{i+1} &= X^{i} + \alpha_i  \cdot G_{X^i},
\end{align}
where $\alpha_i$ is the step size at the $i$-th iteration, and $X^0=Y$. Assume the optimization convergences at $N$-th iteration, we then obtain the restored image $\hat{X}$ as $\hat{X}=X^N$, which is an estimate of the lesion-free version of the observed image $Y$. As we assumed an additive model, the estimated outlier lesion can be revealed as $\hat{D} = Y-\hat{X}$, providing a pixel-wise detection of the lesion. Taking both hypo-intense and hyper-intense areas into account, the absolute value of the difference can also be used as the final detected lesion, as $|\hat{D}|=|Y-\hat{X}|$.

\subsubsection{Data consistency}
\label{DataConsistency}
The likelihood $P(Y|X)$ is a critical component in the MAP estimation. 
As explained in the previous section, it reflects the modeling assumptions on $D$ through how it measures the deviation between $X$ and $Y$. 
Penalizing certain type of deviations more than the others, the likelihood term encodes a preference over $D$ and, thus influence the entire MAP estimation. 
In the unsupervised detection setting, we are aiming for detecting any type of lesions in contrast to the supervised detection setting, where the goal is to detect specific type of lesions. 
As a result, $P(Y|X)$ should not be based on lesion-specific information, such as intensity or specific shape features. 
Generic and mathematically driven likelihood terms, instead of data-driven likelihood terms, can provide the generality required for unsupervised detection.

In this work, we adopt a likelihood term that prefers $D$ to be composed of larger, continuous blocks over many isolated islands. 
Total Variation (TV) norm~\cite{rudin1992nonlinear} formulates this preference and can be easily used as the likelihood term in the MAP estimation. 
Using TV norm, we define the final restoration problem as

\begin{equation}
    \label{eqn:final}\hat{X} = \arg\max_X \left[-\lambda||X - Y||_{TV} + \ELBO(X)\right],\ \lambda>0.
\end{equation}
The TV norm is weighted by $\lambda$, which balances the magnitudes of gradients from the two terms during the gradient ascent optimization. 
An effective weight prevents large deviation between $X$ and $Y$, and results in accurate detection performance.  
Determining $\lambda$ weight is not trivial in the most generic setting. 
Instead, we propose a heuristic method to tune the weight using images from healthy individuals. 

\subsubsection{Determining the weight parameter $\lambda$}
\label{sec:lambda_selection}
Intuitively, images from healthy individuals are lesion-free and, ideally, the detection method should not detect lesions in them. 
Based on this intuition, we determine $\lambda$ that minimizes the change caused by the MAP estimation on healthy images. 
We measure the change caused by the restoration using the $\ell_1$ distance between the restored image and the input images for a small validation set composed only of healthy images,
\begin{equation}
\epsilon(\lambda) = \frac{1}{S}\sum_{s}\left|Y_s - \hat{X}_{\lambda,s}\right|,
\end{equation}
where we denote dependence of the restoration to $\lambda$ with the subscript and $S$ denotes the number of validation images used for the measurement. 
We perform a parameter search in a wide range of possible values to determine the smallest $\lambda$ that yields the smallest $\epsilon(\lambda)$. 

The described heuristic approach relies on the fact that even lesion-free images will be changed during the restoration. 
This happens if the $\ELBO$ term yields non-zero gradients for images composed only of healthy anatomy. There are three reasons why this can happen: 1) $\log P(X)$ as modeled with a latent variable model is not a perfect approximation, 2) $\ELBO$ is an approximation to the $\log P(X)$, and 3) $P(X)$ may assign higher probability to certain anatomical formations and appearance. Most likely, all these reasons are affecting the restoration simultaneously, and even healthy images incur change during restoration. As a result, in lesion-free images, we expect very low $\lambda$ values to yield large number of changes in the images. Moreover, very large $\lambda$ values may also yield changes that have low TV norm and slightly higher $\ELBO$. In our experiments, this is indeed what we observed empirically. Examples of the parameter search are plotted in Figure~\ref{fig:lambda_auc} and~\ref{fig:atlas_lambda_auc} showing an optimal $\lambda$ value that minimizes $\epsilon(\lambda)$ for different prior models and datasets. More details are given in Section~\ref{sec:experiments}. 

\subsubsection{Binary detections by thresholding difference maps}\label{sec:thresholds}
The described MAP estimation will yield a restored image $\hat{X}$ and the outlier area as the difference map $\hat{D}$, which is a pixel-wise continuous map. 
To obtain a pixel-wise \emph{lesion segmentation}, which is a binary map, we need to determine a threshold $T$ and label pixels with differences larger than $T$ as lesion and others as normal anatomy. 
Similar to the case described in the previous section, we need to find the threshold using only healthy images in the unsupervised setting. 
Once again, we assume that healthy images show only normal anatomy and the detection method should not find any lesion in a set of validation healthy images. 
The naive approach of setting the threshold to the maximal pixel-wise difference in the healthy images may yield very conservative method with high specificity but low sensitivity. 
Instead, we adopt the approach as in~\cite{konukoglu2018reconstructing} that implements a compromise by setting a limit to the permitted False Positive Rate (FPR) in the detection results. 
Any detection in the lesion-free healthy images will be a false positive. 
The method proposed in \cite{konukoglu2018reconstructing} uses detections in lesion-free healthy images to estimate the FPR for any threshold, and determines the smallest threshold that satisfies a user defined FPR limit.
This FPR limit dependent threshold can then be used to convert the difference maps $\hat{D}$ into binary segmentations by determining the pixels with $|\hat{D}| > T$. 

To test the effectiveness of this approach, we additionally experimented with determining the threshold using the ROC curve as a baseline.
In this approach, the threshold can be determined by finding the pixel-wise difference value that maximizes the difference between true positive and false positive rates, i.e. $TPR-FPR$, on the ROC curve. 
Note that this selection method makes use of ground-truth lesion annotation to obtain the ROC curve using the difference maps between detections and lesion annotations, therefore, it is not unsupervised. 
Nonetheless, this baseline gives optimistic detection results to compare to the results using FPR-based threshold selection. 

\section{Experiments}\label{sec:experiments}
\subsection{Datasets \& Preprocessing}
We used three different MRI datasets, one for training the prior normative distribution and determining the hyper-parameters of the model, i.e. $\lambda$ and $T_{ls}$, and the other two for evaluating the proposed approach for detecting lesions. 

\noindent\textbf{CamCAN\footnote{\url{http://www.cam-can.org/}}}: Cambridge Centre for Ageing and Neuroscience dataset, described in~\cite{taylor2017cambridge}, contains T1-weighted (CamCANT1) and T2-weighted (CamCANT2) brain scans of 652 healthy subjects of ages ranging from 18 to 87, among which 600 subjects were randomly selected as the data for training the prior models VAE and GMVAE, and 52 as the validation data for determining the hyper-parameters. 

\noindent\textbf{BRATS17\footnote{\url{https://www.med.upenn.edu/sbia/brats2018.html}}}: Multimodal Brain Tumor Image Segmentation Challenge dataset, described in~\cite{menze2015multimodal}, contains T1-weighted and T2-weighted brain scans of 285 subjects with brain tumors. Among them, 210 subjects show high-grade glioblastomas and 75 subjects show low-grade gliomas. We downloaded the version of the dataset published in 2017 and only used the T2-weighted images in this dataset, where the lesions appear as hyper-intensity areas. For all the subjects in this dataset, ground truth segmentations of the tumors visible in the T2-weighted images are also available. 

\noindent\textbf{ATLAS\footnote{\url{http://fcon_1000.projects.nitrc.org/indi/retro/atlas.html}}}: Anatomical Tracings of Lesions After Stroke (ATLAS) dataset~, described in~\cite{liew2018large}, contains 220 T1-weighted brain scans of stroke patients. In these images, the lesions appear as hypo-intensity areas. Similar to the BRATS17 dataset, this dataset also has the ground truth pixel-wise segmentations available. 

We trained two sets of prior models, one using T2-weighted and the other using T1-weighted images in the CamCAN datasets. The former is used for the evaluations on the BRATS17 dataset and the latter for ATLAS dataset. All datasets were preprocessed before training. 
We performed skull stripping to compute brain masks and remove the skulls, registered the images to MNI space, histogram matching among the CamCAN subjects using the method proposed in~\cite{nyul2000new}, and normalized pixel intensities for each subject by scaling them as $(I - \mu{(I)})/(\sigma{(I)})$, with $I$ indicating the intensity values and $\mu(I)$ and $\sigma(I)$ were computed on a randomly selected subject and then fixed for the other subjects, the background pixel intensities were set to $-3.5$. Especially for histogram matching, the choice of the reference subject that the other subjects are matched to will not significantly affect the detection results. 
To mitigate possible domain gaps caused by different acquisition protocols between CamCANT2 and BRATS17 and between CamCANT1 and ATLAS, we matched histograms of the BRATS17 subjects to the same CamCAN subject's T2-weighted image as for CamCANT2 and ATLAS subjects to the same CamCAN subject's T1-weighted image as for CamCANT1, both after normalizing the CamCAN subjects among themselves for each modality respectively. BRATS17 and ATLAS are normalized in the same way as CamCAN.

All computations were done in 2D using transversal slices. 
For the sake of computation efficiency, we excluded from the training data the slices in the beginning and end of a scan in the CamCAN datasets, where no brains structures were present. 
We also removed excessive background in all scans of CamCAN, BRATS17 and ATLAS datasets to obtain transversal slices of size $200 \times 200$. 
During testing, detection was applied to all the slices of a test subject independently while the evaluation metrics were computed subject-wise. 

\subsection{Implementation Details}
We trained two sets of normative prior models using the VAE and GMVAE respectively. For each method, a prior model was trained for T1-weighted images and another model for T2-weighted images using the CamCAN dataset. 
The BRATS17 and ATLAS images were not used during the training of the prior models. 

\begin{figure}[ !ht]
    \centering
    \includegraphics[scale=0.4]{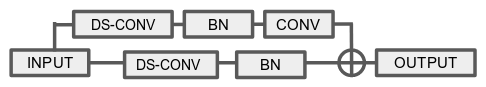}
    \caption{Down-sampling residual block. CONV: convolutional layer, BN: batch normalization, DS-CONV: down-sampling convolution layer. The up-sampling block replaces the down-sampling convolutional layer with up-sampling convolutional layer. }
    \label{fig:upsample_block}
\end{figure}

Our VAE architecture consists of an encoder network and a decoder network. The encoder consists of 6 down-sampling residual blocks (illustrated in Fig.\ref{fig:upsample_block}) with 16, 32, 64, 128, 256 and 512 filters. Down-sampling is done by taking a stride of 2 and up-sampling is done by bilinear up-sampling. The decoder network is symmetrical to the encoder network. The latent variable has a size of $2\times 2\times 512$. We implemented GMVAE following the repository \cite{dilokthanakul2016deep}\footnote{\url{https://github.com/Nat-D/GMVAE}}, except that the GMVAE shared the same network structures for the encoder and decoder as the VAE model to ensure a fair comparison between the two models. The latent space of GMAVE was therefore of the same dimension as the VAE. To extensively evaluate the model performance, we also trained the GMVAE model with different numbers of Gaussian mixtures and different latent dimensions, see Section \ref{sec:anal_gm_prior}. \verb|leaky-relu| activation is applied to all hidden layers while \verb|identity| activation is applied only to output layers and layers that connect to the latent variables. 

To provide the baseline achieved by supervised methods, a U-Net is implemented with the same encoder and decoder structure while the encoder and decoder are connected by skip connections. Cross-entropy loss is used to train the network. The choice of the U-Net structure is to ensure consistency among experiments and may not over-perform the state-of-the-art accuracy on the BRATS17 leaderboard.  

We applied the proposed MAP-based detection method to the test images using the appropriate prior models. 
By observing the convergence of gradient ascent optimization, we performed 500 steps of restoration, with a step size of $5\times 10^{-3}$ in the first 100 iterations and $3\times 10^{-3}$ in the following iterations. 
This training strategy was observed to lead to higher MAP values at the end of the optimization, evaluation metrics were not used to determine the strategy. We refer to our proposed versions as VAE (TV) and GMVAE (TV) while presenting the results.

For model comparison, we implemented four different methods that uses prior projection approach explained in the introduction. These are VAE-256, VAE-128, AAE-128, which corresponding to models trained with images resized to 256$\times$256 and 128$\times$128, as described in \cite{chen2018deep}, and AnoGAN proposed by \cite{schlegl2017unsupervised}, with all hyper-paremeters tuned with our datasets. 
We compared the detection accuracy of the proposed MAP-based restoration approach using VAE and GMVAE to these prior works.

Lastly, we used different evaluation metrics to present a better picture of the performance of the proposed model and the ones we used as comparison. First, we computed the ROC curves using the continuous valued detections $\hat{D}$. ROC curves were computed using the detections in the entire dataset. From the ROC curve, we extracted the Area-Under-the-Curve (AUC), which was used as the first metric. 

Using the procedure explained in Section~\ref{sec:thresholds} we also computed thresholds at different FPR limits for all the algorithms, at 1\%, 5\% and 10\%. We thresholded the absolute valued detection maps, e.g. $\hat{D}$ for our model, to construct binary detection maps and computed the Dice's Similarity Coefficient between the ground truth segmentations and the detections, referred to as DSC1, DSC5 and DSC10, respectively. The DSC values were computed subject-wise, pooling all the transversal slices of the subject together. To understand the efficacy of the threshold determining method, we also extracted the threshold from the ROC curve that yielded the biggest difference between the true positive and false positive rates, as explained previously. This threshold was also used to compute the DSC for all the methods and forms a baseline, and we refer to it as DSC\_AUC while presenting the results. 

\begin{table*}[t]
    \caption{Quantitative results on the BRATS17 dataset: Summarized AUC and DSC values for the proposed MAP-based detection with VAE and GMVAE as prior models, and other baseline methods are presented in the table. DSC1, DSC5, DSC10 are DSC values calculated with automatically determined thresholds that corresponds to limiting FPR to 1\%, 5\% and 10\% on the training set, respectively. For GMVAE (TV) the automatically determined weighting parameter $\lambda$ was 5.0 for $c=3$, 4.8 for $c=6$ and 4.0 for $c=9$ . For VAE (TV) this value was 2.4. *shows the DICE score for U-Net. The best results are indicated in bold, higher results are better. }
    \begin{center}
    \begin{tabular}{ p{4cm} p{1 cm}p{2.2cm} p{2.2cm} p{2.2cm} p{2.2cm}}
        \toprule
        Methods  & AUC  & DSC1  & DSC5   & DSC10 & DSC\_AUC \\
        \midrule
        VAE (TV) (ours) & 0.80 & \textbf{0.34$\pm$0.20} & 0.36$\pm$0.27 & 0.40$\pm$0.24 & 0.34$\pm$0.18 \\
        GMVAE (TV), c=3 (ours) & 0.82 & 0.21$\pm$0.20 & 0.39$\pm$0.22 & 0.38$\pm$0.20 & 0.35$\pm$0.20 \\
        GMVAE (TV), c=6 (ours) & 0.81 & 0.31$\pm$0.14 & 0.40$\pm$0.22 & 0.37$\pm$0.16 & 0.33$\pm$0.19\\
        GMVAE (TV), c=9 (ours) & \textbf{0.83} & 0.32$\pm$0.23 & \textbf{0.45$\pm$0.20} & \textbf{0.42$\pm$0.19} & \textbf{0.36$\pm$0.19} \\
        \midrule
        VAE-256 & 0.67 &  0.06$\pm$0.06 & 0.18$\pm$0.13  & 0.25$\pm$0.20 &  0.20$\pm$0.14  \\
        VAE-128 & 0.69  &  0.09$\pm$0.06 & 0.19$\pm$0.15 & 0.26$\pm$0.17& 0.22$\pm$0.14 \\        AAE-128  &0.70   &  0.03$\pm$0.03 & 0.18$\pm$0.14 & 0.23$\pm$0.15& 0.23$\pm$0.13 \\
        AnoGAN  & 0.65 & 0.02$\pm$0.02 &0.10$\pm$0.06 & 0.19$\pm0.13$& 0.19$\pm$0.10 \\
        \midrule
        U-Net (supervised) & / & / & / & / & 0.85*\\
        \bottomrule    \end{tabular}
    \label{tab:brats_table}
    \end{center}
\end{table*}

\begin{table*}[t]
    \caption{Qualitative results on the ATLAS dataset. Summarized AUC and DSC for GMVAE(TV) and baseline methods. Same metrics as in Table~\ref{tab:brats_table} are presented in this table. For the ATLAS dataset, for GMVAE (TV) the automatically determined $\lambda$ was 3.0 for $c=3$, 4.0 for $c=6$ and 6.0 for $c=9$. For VAE (TV) this value was 3.0. *shows the DICE score for U-Net. The best results are indicated in bold, higher results are better.}
    \begin{center}
    \begin{tabular}{  p{4cm}  p{1 cm}  p{2.2cm}  p{2.2cm}  p{2.2cm}  p{2.2cm}}
        \toprule
        Methods        &AUC    & DSC1          & DSC5         & DSC10& DSC\_AUC\\\hline
        VAE(TV) (ours)  & \textbf{0.79} & \textbf{0.10$\pm$0.06} & 0.11$\pm$0.05  & \textbf{0.11$\pm$0.05} & \textbf{0.11$\pm$0.07}\\
        GMVAE(TV), c=3 (ours) & \textbf{0.79} & 0.06$\pm$0.06 & 0.09$\pm$0.07 & 0.08$\pm$0.07  & 0.07$\pm$0.07 \\
        GMVAE(TV), c=6 (ours) & \textbf{0.79} & 0.10$\pm$0.09 & \textbf{0.12$\pm$0.12} & 0.08$\pm$0.07 & 0.08$\pm$0.08  \\
        GMVAE(TV), c=9 (ours) & 0.77 & 0.08$\pm$0.07 & 0.10$\pm$0.08 & 0.07$\pm$0.07  & 0.08$\pm$0.07 \\\midrule
        VAE-256  & 0.66 &  0.00$\pm$0.00 & 0.01$\pm$0.01 & 0.02$\pm$0.02 & 0.02$\pm$0.02\\
        VAE-128   & 0.64 &  0.00$\pm$0.00 & 0.01$\pm$0.01 & 0.01$\pm$0.01 & 0.01$\pm$0.01\\
        AAE-128  &  0.63 & 0.00$\pm$0.00 & 0.01$\pm$0.01 & 0.01$\pm$0.01 & 0.01$\pm$0.01\\
        AnoGAN  & 0.64 &  0.00$\pm$0.00 & 0.01$\pm$0.01 & 0.02$\pm$0.02 & 0.01$\pm$0.01\\
        \midrule
        U-Net (supervised) & / & / & / & / & 0.50*\\
        \bottomrule
    \end{tabular}
    \end{center}
    \label{tab:atlas_table}

\end{table*}

\begin{figure*}[ !ht]
    \begin{subfigure}[t]{0.5\textwidth}
    \label{fig:roc_brats}%
    \centering
    \includegraphics[width=1.\textwidth]{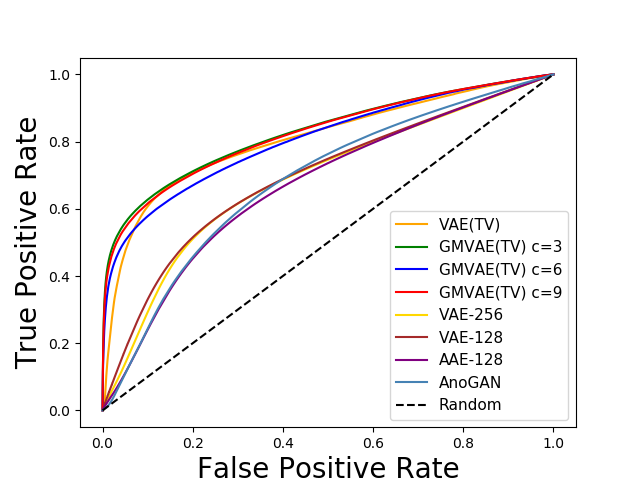}    
    
    \end{subfigure}
    \begin{subfigure}[t]{0.5\textwidth}
    \label{fig:roc_atlas}%
        \centering
    \includegraphics[width=1.\textwidth]{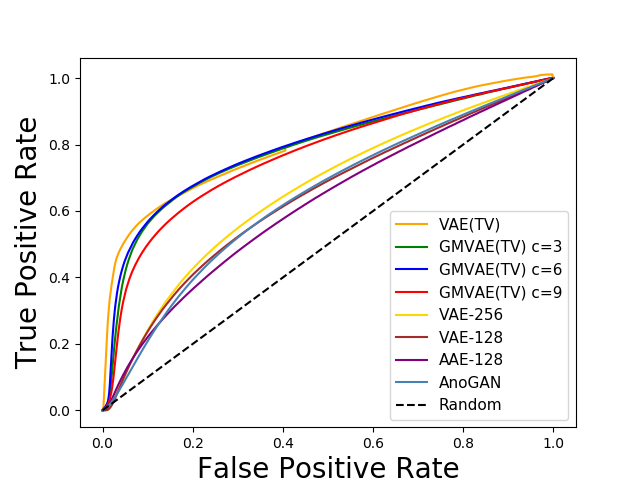}
    
    \end{subfigure}
    \caption{ROC curves of proposed methods against baselines. Left: ROC curves of all methods computed on the BRATS17 dataset. Right: ROC curves of all methods computed on the ATLAS dataset. ROC curves were computed with the entire datasets. Dashed line indicates the random detection.}
    \label{fig:rocs}
\end{figure*}

\begin{figure*}[ !httb]
    \centering
    \includegraphics[scale=0.35]{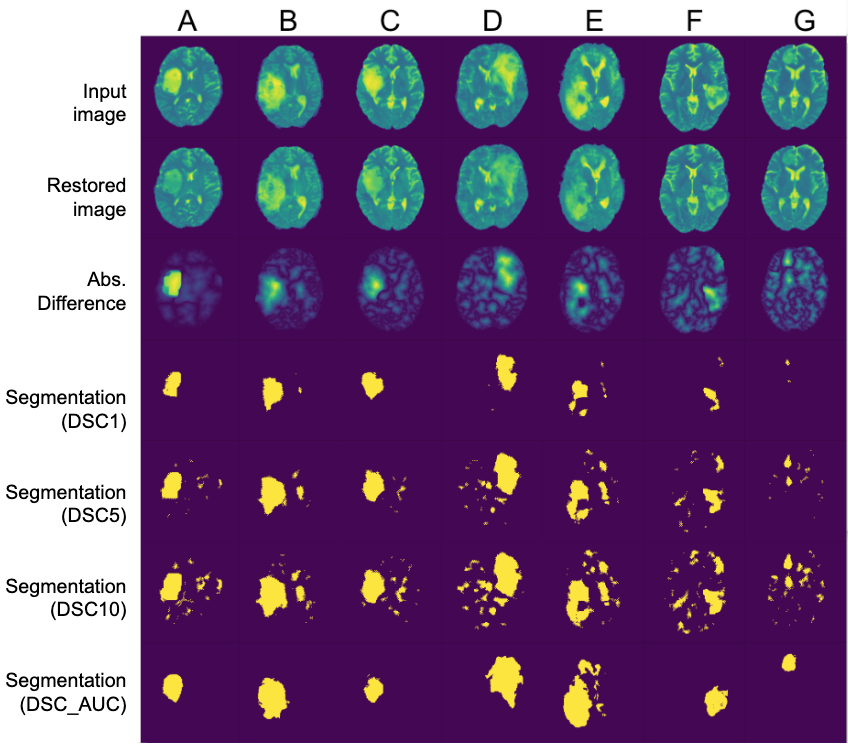}
    \caption{Visual examples of lesion detection with the GMVAE (TV) model with $c=9$ for the BRATS17 dataset. Rows 1 to 7 show the input images, restored images, continuous detection maps $\hat{D}$, binary detection maps based on automatically determined thresholds at \%1, \%5 and \%10 FPR limits, and ground-truth lesion segmentation, respectively. Columns A to G are images of 7 randomly selected subjects from the dataset.}
    \label{fig:brats_visualization}
\end{figure*}

\begin{figure*}[ !httb]
    \centering
    \includegraphics[scale=0.4]{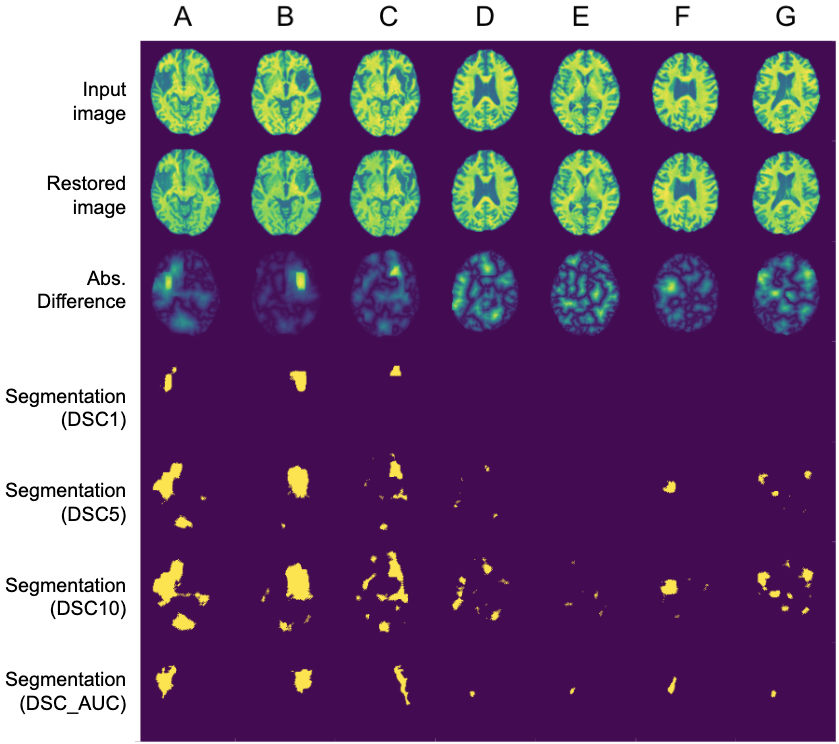}
    \caption{Lesion detection with GMVAE(TV) with $c=6$ for ATLAS dataset. Visualization is the same as Figure~\ref{fig:brats_visualization}.}
    \label{fig:atlas_visualization}
\end{figure*}

\subsection{Results}
\subsubsection{Model Performance and Comparisons}
We present the quantitative results in Tables~\ref{tab:brats_table} and~\ref{tab:atlas_table} with summary metrics. The tables present a single AUC value for each method as this metric was computed over the entire datasets, while DSC values were computed subject-wise and the table presents mean and standard-deviations. The two variations of the proposed MAP-based detection method, with VAE and GMVAE as the prior models, are shown in the top rows. For GMVAE, we show results for different number of clusters, i.e. $c=3, 6 \text{ and }9$. The baseline methods are shown in the following rows. 

In our experiments with the BRATS17 dataset, the baseline methods, namely VAE-256, VAE-128, AAE-128 and AnoGAN, yielded values of 0.70 or lower. In comparison, the proposed method improved the AUC score to over 0.80 for all the prior models. 
Similar results were obtained in the experiments with the ATLAS dataset. 
The variations of the proposed MAP-based detection method yielded higher AUC for both prior terms. 
The GMVAE prior model yielded a modestly higher AUC than the VAE prior model in the BRATS17 dataset and the same AUC values in the Atlas dataset. 
The ROC curves for both the datasets are presented in Figure~\ref{fig:rocs}. 


The DSC values revealed a similar picture as the AUC with respect to baseline methods.
In the experiments with the BRATS17 dataset, in a setting with conservative FPR limit, i.e. 1\%, baseline methods achieved DSC values lower than 0.10. 
The proposed method improved the DSC to 0.34 with VAE and more than 0.20 for all the GMVAE variations on average.
When the FPR limit was increased to 5\%, a less conservative limit, all the methods yielded higher DSC than in the 1\% limit setting. 
Despite the increase, the DSC of the baseline methods remained lower than 0.20.
The proposed method improved mean DSC substantially for all the prior terms over the best baseline methods. 
In the least conservative setting with FPR limit at 10\%, DSC value of the best baseline method (VAE-128) increased to 0.26, which is comparable to the DSC the proposed method with GMVAE at 1\% FPR limit. 
The DSC values of the other baseline methods were also higher at this FPR limit. 

Similar trends were also observed in the experiments with the ATLAS dataset. 
DSC values of the proposed methods were substantially higher than the ones yielded by baseline methods for both prior terms.
However, the DSC values were substantially lower for all the methods in the ATLAS dataset compared to those in the BRATS17 dataset. 
The lower DSC values are due to the difficulty of the detection task in the ATLAS dataset. 
As can be seen in the visual examples shown in Figure~\ref{fig:atlas_visualization}, lesions can be smaller and have similar intensities as the normal structures. 
These lower DSC values also demonstrate the current limitations in the performance of the unsupervised lesion detection approaches, and suggests substantial room for improvement.

In our experiments, there was not a clear winner between VAE and GMVAE priors in the MAP restoration approach.
GMVAE prior with $c=9$ yielded higher AUC and mean DSC at 5\% and 10\% FPR limits than using VAE prior on the BRATS17 dataset. 
However, the increase was not observed in the ATLAS dataset. 

The DSC\_AUC values are presented at the left most column in both the tables. These scores were obtained with thresholds computed retrospectively with the knowledge of the ROC curves, therefore, they cannot be used for evaluation. However, they are useful for evaluating the efficacy of the method for identifying the thresholds automatically, which was described in Section~\ref{sec:thresholds}. We observe that DSC values obtained with the automatic method were similar to those set using ROC curves for all the methods, suggesting that the automatic method is indeed appropriate for identifying thresholds for generating binary detection maps. 

\begin{figure*}[ !httb]
    \begin{subfigure}[b]{.5\textwidth}{
    \includegraphics[scale=0.5]{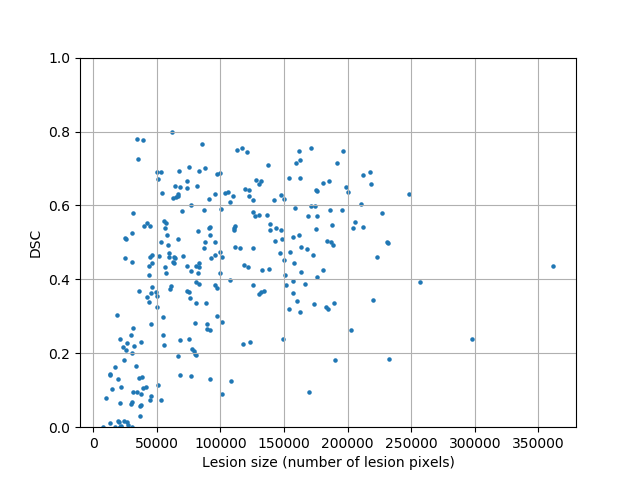}
    \caption{\label{size_dsc_brats}}}
    \end{subfigure}
    \begin{subfigure}[b]{.5\textwidth}{
    \includegraphics[scale=0.5]{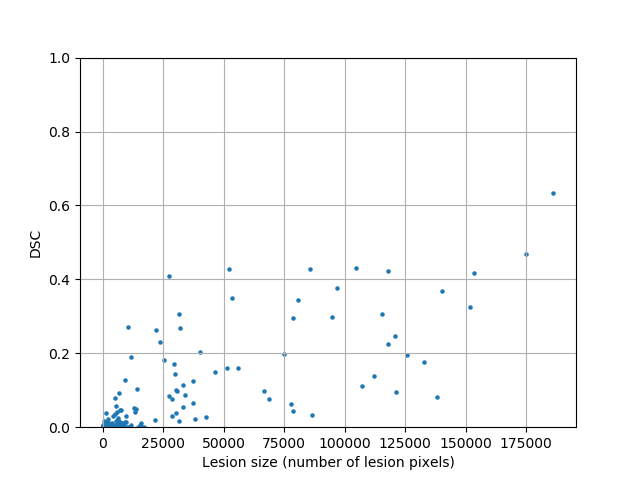}
    \caption{\label{size_dsc_atlas}}}
    \end{subfigure}

\caption{Lesion size vs. Accuracy. The accuracy is measured as DSC5 and plotted on the y-axis, and the lesion size is measured as the number of annotated pixels for the lesion and is plotted on the x-axis. The model and parameters that achieve the highest DSC on each dataset are used. (a) BRATS17 dataset, results obtained by the trained GMVAE(TV, c=9). (b) ATLAS dataset, results obtained by the trained GMVAE(TV, c=6).}
\end{figure*}

\begin{figure*}[ !httb]
    \begin{subfigure}[b]{.5\textwidth}{
    \includegraphics[scale=0.5]{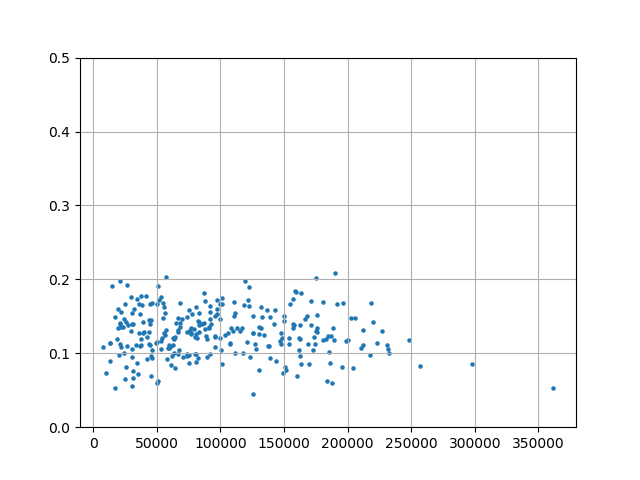}
    \caption{\label{size_brats}}}
    \end{subfigure}
    \begin{subfigure}[b]{.5\textwidth}{
    \includegraphics[scale=0.5]{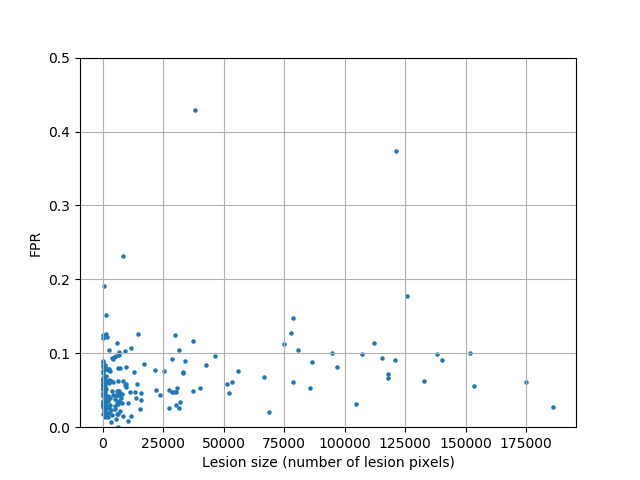}
    \caption{\label{size_atlas}}}
    \end{subfigure}

\caption{Lesion size vs. FPR. FPR is computed for each subject with the threshold $\hat{D}$ and ground truth lesion segmentation and is plotted on the y-axis, and the lesion size is measured as the number of annotated pixels for the lesion and is plotted on the x-axis. The model and parameters that achieve the highest DSC on each dataset are used. (a) BRATS17 dataset, results obtained by the trained GMVAE(TV, c=9). (b) ATLAS dataset, results obtained by the trained GMVAE(TV, c=6).}
\end{figure*}

\begin{figure*}

    \begin{subfigure}[b]{.5\textwidth}
    \centering
    \includegraphics[scale=0.32]{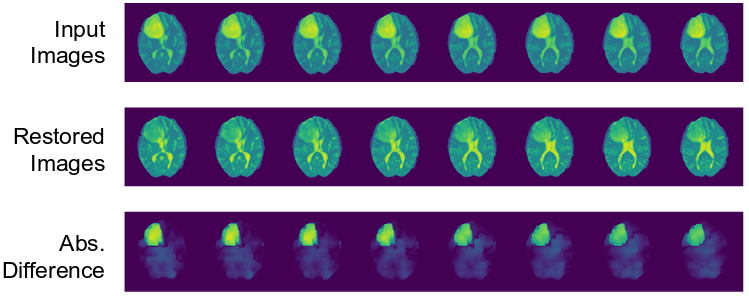}
    \caption{\label{fig:brats_3d_consistency}}
    \end{subfigure}
    \begin{subfigure}[b]{.5\textwidth}
    \centering
    \includegraphics[scale=0.32]{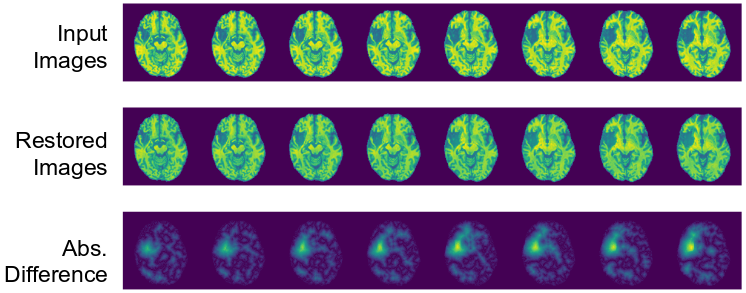}
    \caption{\label{fig:atlas_3d_consistency}}
    \end{subfigure}
    \caption{Detection on consecutive slices on (a) slices 72-80 from subject Brats17\_TCIA\_490\_1 from BRATS17, prior model trained with GMVAE (TV, c=9) and (b) slices 55-63 from subject 031859\_t1w\_deface\_stx from ATLAS, prior model trained with GMVAE (TV, c=6). We notice that the visualised ATLAS images are horizontally flipped, and yet this does not affect training or evaluation. }
\end{figure*}

Lastly, we present visual results obtained by the proposed method using the GMVAE prior model in Figures~\ref{fig:brats_visualization} and~\ref{fig:atlas_visualization}. 
Figures show both continuous valued detections, $\hat{D}$, and binary detection maps thresholded at the selected FPR limits. 
From the BRATS17 dataset, to give a better perspective of the performance, we selected images with large and small tumors, and show how the detection performance was affected by tumor sizes. 
Cases with large tumors are shown in columns A to D and small tumors in columns E to G in Figure~\ref{fig:brats_visualization}. 
In the restored images (row 2), the abnormal intensities have been substantially reduced to a normal range, resulting in detectable intensity changes in the difference images (row 3). 
For tumors of large sizes, the thresholded detection results matches well with the ground truth segmentations. 
Smaller tumors were more difficult for the method to detect accurately. Particularly, the method may include "abnormal-looking" healthy tissues in the detection in addition to the small tumors (column F) or produce no detections, overlooking the tumors.
We also note that although the detection becomes less accurate on small tumors, the method does not detect healthy pixels in a large region to be abnormal. 
Comparing binary detections at different FPR limits, we observe that as the FPR limit is higher the detections include more tumor pixels as well as non-tumor pixels, as expected. 

Visual results for the ATLAS dataset are shown in Figure~\ref{fig:atlas_visualization}. This problem is much harder due to the size and intensity characteristics of the lesions. Comparing the continuous valued detection maps shown in the third row and the ground truth segmentations in the last row shows that the model is able to highlight the lesions but with additional areas including normal structures. Binary detections presented in the fourth-to-sixth rows show this more clearly. While the lesion pixels are often captured in the outlier maps, multiple areas displaying normal anatomy are also captured, yielding lower DSC values at the end. 

\subsection{Accuracy Analysis with Lesion Size}
To observe the relation between detection accuracy and tumor size, we calculated the dice scores for all test subjects in BRATS17 and ATLAS respectively. Specifically, we measured the lesion size as the annotated lesion pixels. The relation is visualized as scatter-plots in Figure 5 using the best-performing model for the two datasets. 

The scatter-plots for the two dataset both indicate a tendency that higher dice scores are often obtained on subjects with larger lesions. The tendency is more obvious on ATLAS dataset while it is weak on BRATS17. As shown as Figure 6(a) ad (b), the FPR for BRATS17 is mostly between 0.10 and 0.20 while the FPR for ATLAS is mostly lower than 0.10. Although the accuracy on BRATS is significantly higher than ATLAS, the low FPR on ATLAS may be caused by the large amount of true negatives. On the other hand, as in Figure 5(b), the higher dice score achieved on ATLAS is for a subject with the largest lesion and very small lesions with size $<$ 25000 pixels mostly give dice scores lower than the best average dice of 0.12. However, in Figure 5(a),this accuracy-size relation is less obvious on BRATS17 than ATLAS. With the size larger than approximately 25000, dice scores higher than the optimal average dice, 0.45, can be achieved. As the lesion size increases, the dice scores can vary from 0.1 to 0.8. The highest dice score is also not obtained for the relatively larger lesions with more than 200,000 pixels. Besides the lesion size, other characteristics of the lesion, such as intensity, location and shape, may also affect the accuracy in a complicated way.


\subsection{3D consistency in detection}
As the scans come as 3D volumes, we would like to observe if the purposed model has detection consistency in the 3D manner. As shown in Figure \ref{fig:brats_3d_consistency} and \ref{fig:atlas_3d_consistency}, the detected lesions are consistent across the slices. 

\subsection{Model Analysis with GMVAE prior}
To further analyze the behavior of the method, we investigated the effects of the model hyperparameters on the performance and the convergence behavior of the optimization. 
Our analyses focuses on the proposed method used with GMVAE, as this version lead to slightly higher performance and it also has additional hyperparameters. 
We experimented with data consistency weight $\lambda$, the number of Gaussian mixtures $c$ and the dimension of the latent variable $dim_z$. 

\begin{table*}[t]
    \caption{AUC/DSC values for varying latent space dimension $dim_z$ and number of clusters $c$ in GMVAE (TV). Mean and standard deviations are shown for DSC at different FPR limits. Results are shown for the BRATS17 dataset. Best results are indicated in bold.}
    \begin{center}
    \begin{tabular}{ p{1 cm}  p{2cm}  p{2cm}  p{2cm}  p{2cm}  p{2cm}   p{2cm}}
    
        \toprule
        $c$ & $dim_z$ &AUC  & DSC1 & DSC5  & DSC10& DSC\_AUC\\\midrule
        
        \multirow{ 3}{*}{c=1} & 256  & 0.80 & 0.32$\pm$0.22 & 0.34$\pm$0.24 & 0.36$\pm$0.24 & 0.32$\pm$0.19 \\
        & 512  & 0.80 & \textbf{0.33$\pm$0.24} & 0.36$\pm$0.24 & 0.39$\pm$0.23  & 0.35$\pm$0.19 \\
        & 1024 & 0.79 & 0.30$\pm$0.22 & 0.31$\pm$0.23 & 0.33$\pm$0.23  &0.31$\pm$0.18  \\\midrule
        
        \multirow{ 3}{*}{c=3} & 256  & 0.82 & 0.20$\pm$0.18 & 0.32$\pm$0.23 & 0.30$\pm$0.21 & 0.30$\pm$0.18 \\
        & 512  & 0.82 & 0.21$\pm$0.20 & 0.39$\pm$0.22 & 0.38$\pm$0.20 & 0.35$\pm$0.20 \\
        & 1024 & 0.82 & 0.20$\pm$0.19 & 0.34$\pm$0.19 & 0.33$\pm$0.19 & 0.32$\pm$0.20 \\\midrule
        \multirow{ 3}{*}{c=6} & 256  & 0.81  & 0.30$\pm$0.21 & 0.40$\pm$0.18 & 0.35$\pm$0.20 & 0.34$\pm$0.18  \\
        & 512  & 0.81 &
        0.32$\pm$0.23 & 0.44$\pm$0.21 & 0.41$\pm$0.20 & 0.35$\pm$0.18 \\
        & 1024 & 0.81 & 0.29$\pm$0.22 & 0.38$\pm$0.23 & 0.35$\pm$0.19 & 0.33$\pm$0.19  \\\midrule
        \multirow{ 3}{*}{c=9} & 256  & 0.82 & 0.32$\pm$0.30  & 0.38$\pm$0.19 & 0.36$\pm$0.15 & 0.31$\pm$0.15 \\
        & 512  & \textbf{0.83} & 0.31$\pm$0.23 & \textbf{0.45$\pm$0.20} & \textbf{0.42$\pm$0.19} & \textbf{0.36$\pm$0.19}  \\
        & 1024 & 0.82 & 0.30$\pm$0.14 &
        0.38$\pm$0.22 & 0.35$\pm$0.20 & 0.34$\pm$0.18 \\
        \bottomrule
    \end{tabular}
    \end{center}
    \label{tab:clusters}
\end{table*}

\subsubsection{Data Consistency Weighed with Different $\lambda$ Values}
\begin{figure*}[ !htb]
    \begin{subfigure}[b]{.33\textwidth}
    \includegraphics[height=.72\textwidth]{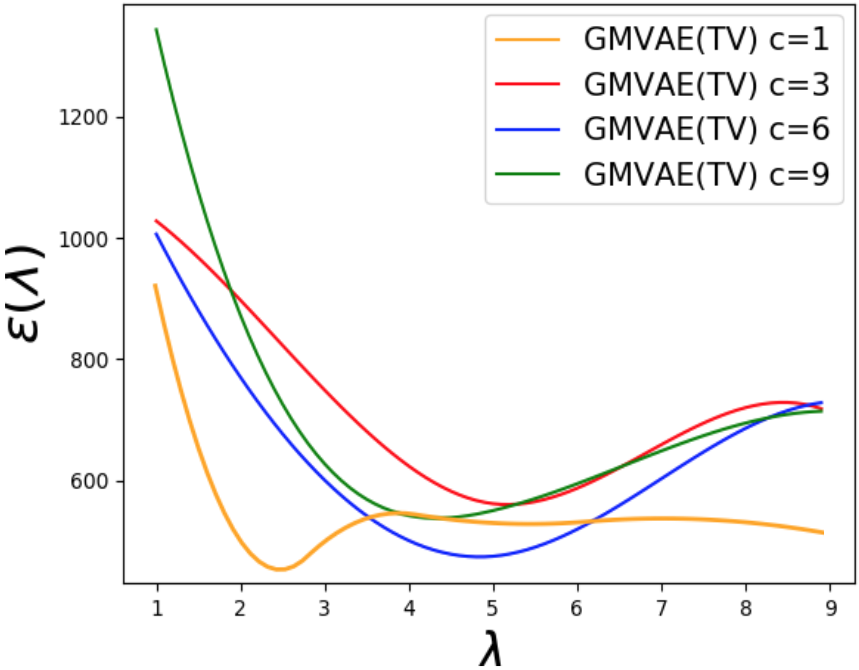}
    \caption{\label{fig:lambda_auc}}
    \end{subfigure}
    \begin{subfigure}[b]{.33\textwidth}
    \includegraphics[height=0.72\textwidth]{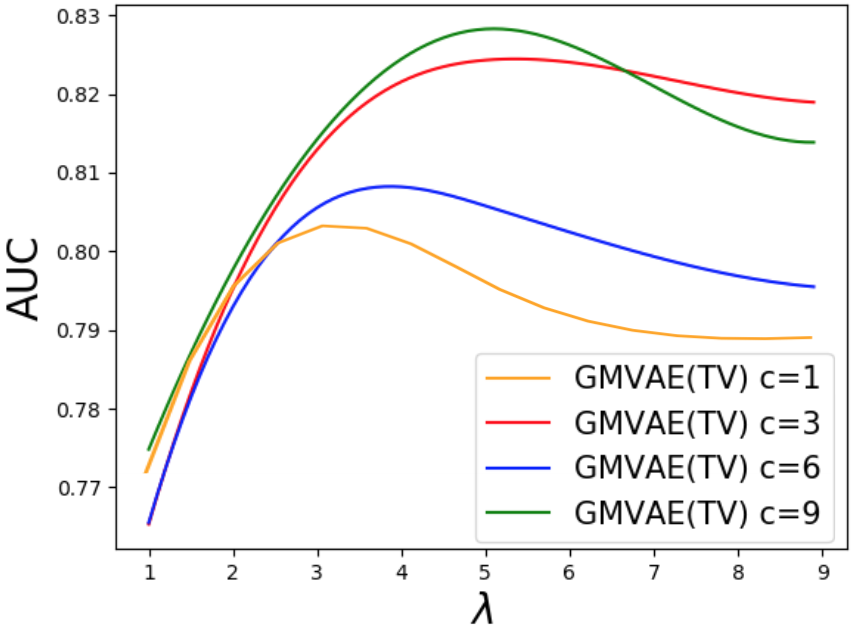}
    \caption{\label{fig:lambda_variation}}
    \end{subfigure}
    \begin{subfigure}[b]{.33\textwidth}
    \includegraphics[height=0.72\textwidth]{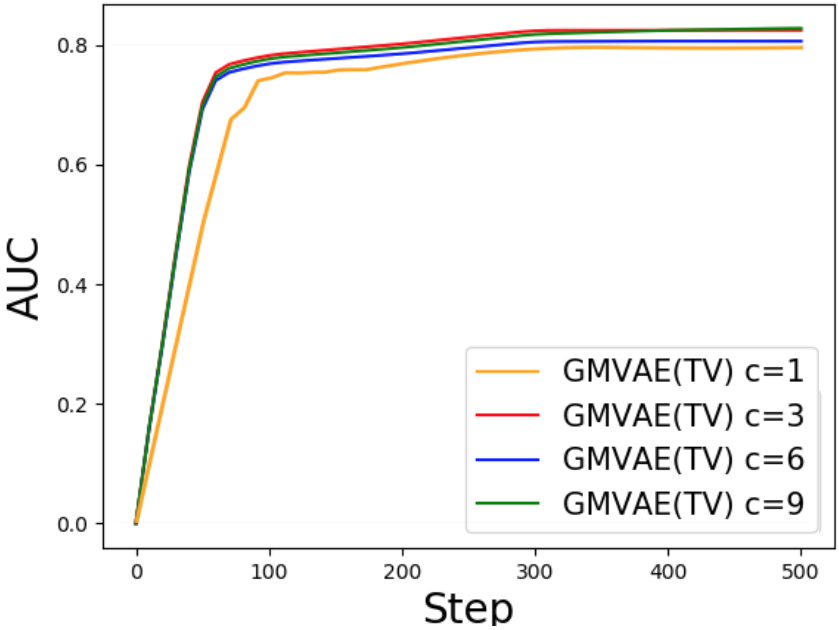}
    \caption{\label{fig:step_auc}}
    \end{subfigure}
     \caption{ \label{fig:lambda_analysis} Analysis graphs obtained using the proposed method with the GMVAE prior trained with T2-weighted images in the CamCANT2 dataset. (a) $\epsilon(\lambda)$ vs. $\lambda$ curve used to determine the optimal hyperparameter $\lambda$ value using the validation images from the CamCANT2 dataset consisting only of normal anatomy. (b) AUC vs. $\lambda$ curve showing stable performance on the BRATS17 dataset over a range of $\lambda$ values. This curve is not used to determine the hyperparameter but only for analysis. (c) Evolution of the AUC over the gradient-ascent iterations during MAP-optimization showing convergence. The AUC values are computed on the BRATS17 dataset.}
     \label{fig:t2_curves}
\end{figure*}

\begin{figure*}[ !httb]
    \begin{subfigure}[b]{.33\textwidth}
    \includegraphics[height=.71\textwidth]{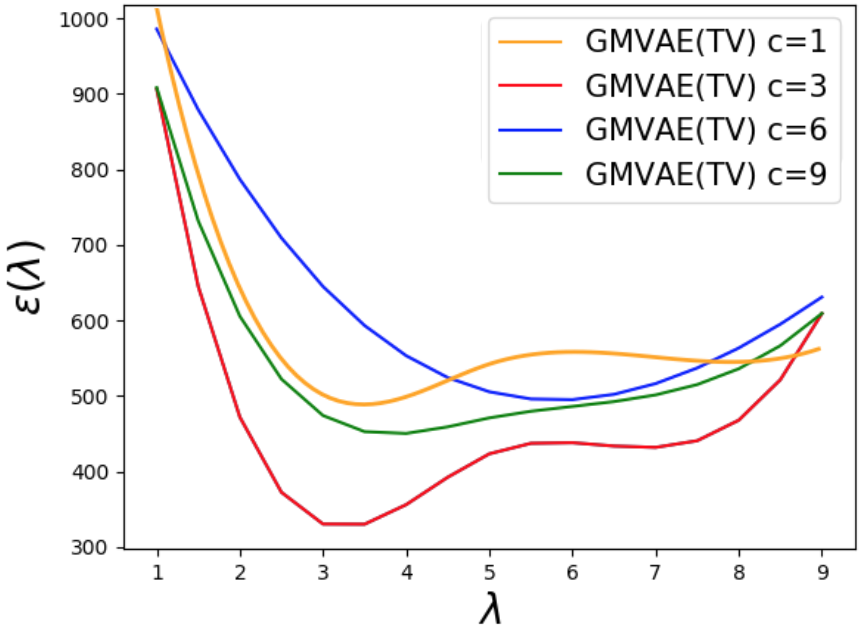}
    \caption{\label{fig:atlas_lambda_auc}}
    \end{subfigure}
    \begin{subfigure}[b]{.33\textwidth}
    \includegraphics[height=0.72\textwidth]{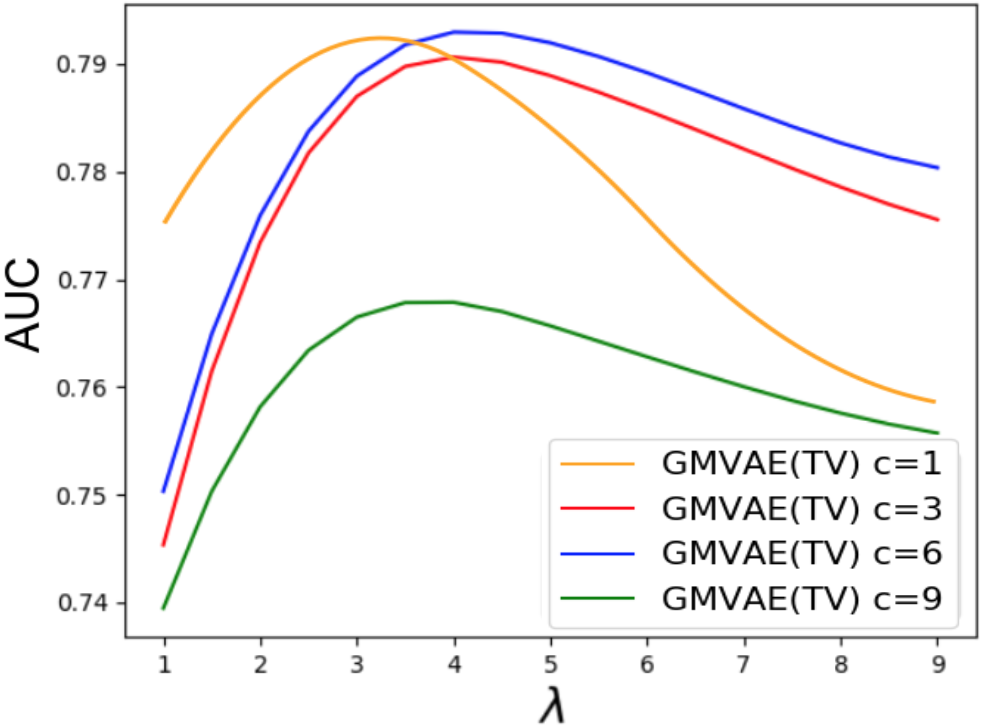}
    \caption{\label{fig:atlas_lambda_variation}}
    \end{subfigure}
    \begin{subfigure}[b]{.33\textwidth}
    \includegraphics[height=0.72\textwidth]{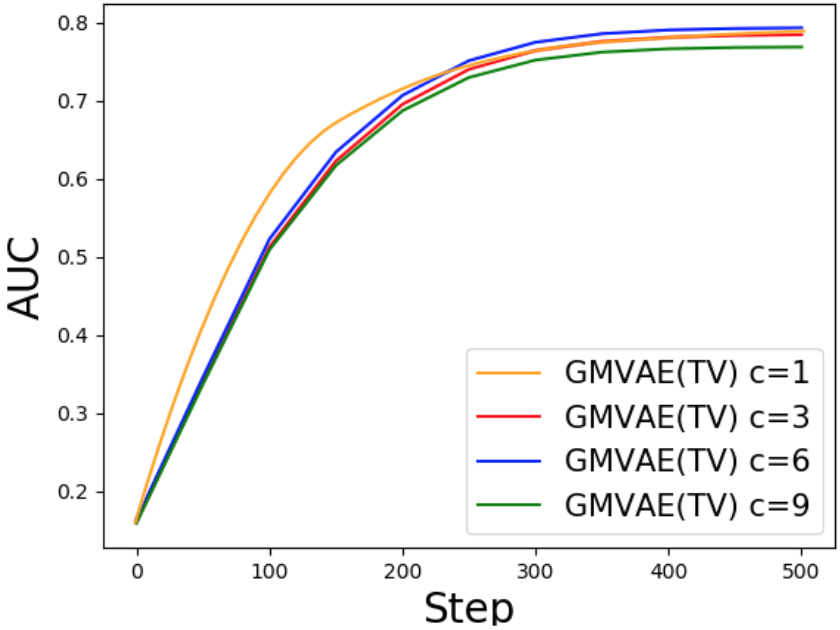}
    \caption{\label{fig:atlas_step_auc}}
    \end{subfigure}
     \caption{ Similar plots are shown for T1-weighted images as in Figure~\ref{fig:lambda_analysis},  (a) $\epsilon(\lambda)$ vs. $\lambda$ curve. (b) AUC vs. $\lambda$ curve. (c) Evolution of the AUC over the gradient-ascent iterations during MAP-optimization showing convergence. GMVAE priors are trained with the T1-weighted images in the CamCANT1 dataset and the detections are evaluated on the ATLAS dataset.}
     \label{fig:t1_curves}
\end{figure*}

The weight parameter $\lambda$ is actually selected automatically using the method described in Section~\ref{sec:lambda_selection}. For this selection, the method relies on the error term $\epsilon(\lambda)$ computed using a set of validation images. Here we investigate how $\epsilon(\lambda)$ changes. First, we plot $\lambda$ vs $\epsilon(\lambda)$ in Figures~\ref{fig:lambda_auc} and~\ref{fig:atlas_lambda_auc} to show the behavior of this term. The plots are generated using GMVAE prior models that was trained with T2- and T1-weighted images from the CamCANT2 dataset, respectively. $\epsilon(\lambda)$ was evaluated using the validation images for different $\lambda$ values. We observe a dip in the curves indicating optimal $\lambda$ values that yielded the least amount of change on validation images consisting only of healthy anatomy for each $c$. These values were used in the experiments on the BRATS17 and ATLAS dataset with the GMVAE prior model presented previously. 

A natural question that arises is whether choosing $\lambda$ according to $\epsilon(\lambda)$ corresponds to choosing the best $\lambda$ according to the detection accuracy on a specific lesion dataset. To answer this question and analyze the sensitivity of the detection results to the $\lambda$ parameter, we repeated the detection experiments with multiple $\lambda$ values in the [1.0, 9.0] range and evaluated the performance of these different models using AUCs. Figures~\ref{fig:lambda_variation} and~\ref{fig:atlas_lambda_variation} plots the results of these experiments. 

First, we observe that $\lambda$ value can influence the AUC. Very low $\lambda$ values yielded lower AUC values in both the datasets. Second, we see that the $\lambda$ values chosen using $\epsilon(\lambda)$ are not very far from the $\lambda$ values that yielded the maximum AUC values for all the prior terms in both the datasets. Choosing the $\lambda$ using the method described in Section~\ref{sec:lambda_selection} led to at most 0.01 less than the maximum AUC values. 

\subsubsection{Analysis of the Gaussian Mixture Prior}
\label{sec:anal_gm_prior}

GMVAE uses a Gaussian mixture model as the prior distribution in the latent space to allow fitting more complex normative distributions. This model has two hyper-parameters: number of Gaussian mixtures $c$ and the dimension of the latent space. In this section, we present analysis showing the effect of these parameters on the detection performance on the BRATS17 dataset. 

For a range of the cluster number $c$ and latent space dimension $dim_z$, we trained multiple prior models and detected lesions on the BRATS17 datasets. In Table~\ref{tab:clusters} we present the detection results for the different ranges.  We notice that in terms of AUC, the parameters have minimal effect. For DSC, 512 latent dimensions seems to provide the highest scores for all the FPR limits and $c$. The numerical differences, however, are small compared to the standard deviations. The biggest difference is arguably between $c=3$ and the others in DSC1 and DSC5. Increasing $c$ seems to improve DSC at these FPR limits. Additionally, GMVAE with $c=1$ is a special case of GMVAE and is similar to VAE. We empirically validate this by reporting the results of GMVAE ($c=1$) in Table \ref{tab:clusters} and Figure \ref{fig:t2_curves} and \ref{fig:t1_curves}. GMVAE(TV) with $c=1$ give very similar results to VAE(TV).




\subsubsection{Convergence of Image Restoration}
As the image restoration is performed in an iterative manner, it is important to show that the restoration stably converges at an optimal point. 
We experimentally show the convergence of the restoration process. We evaluated the AUC using the BRATS17 and ATLAS datasets at each 50 steps during the MAP-optimization and plot the evolution of AUC with respect to gradient-ascent steps in Figures~\ref{fig:step_auc} and~\ref{fig:atlas_step_auc}. 
For the restoration on BRATS17, the plot shows a sharp increase in the first 100 iterations and then increases to stable values until 500$^\textrm{th}$ iterations at around AUC$=$0.8 for all the selected $c$ values. For the restoration on ATLAS, the plot shows similar increases where the AUC values increases in the first 300 iterations and stabilizes around 0.80. The plots indicate that the iterative restoration converged and the model performance stably improved with increasing restoration steps.

\section{Conclusions}
We proposed an unsupervised detection method which restores images using an estimated normative prior for healthy images. Specifically, the image prior was learned with autoencoding-based methods, VAE and GMVAE, and the images were iteratively restored with MAP estimation.
Detection on brain lesion datasets showed that the proposed method achieved better AUC and Dice scores, significantly outperforming the existing methods. Extended model analysis on GMVAE indicated that the prior learned by this model is robust to parameter selection and the model shows stable convergence. Meanwhile, the model has limited capability when applied to detect small and unobvious lesions. On the other hand, some lesions may cause large deformations in the surrounding healthy structures. Although the data consistency term prevents the model from detecting such deformations as abnormal lesions, there is limited guarantee that large deformations will not be detected as abnormalities. We have not observed this in our experiments, however, this remains a limitation. To improve the performance, future works may impose noise assumption that better describes the lesions and
use adaptive thresholding selection. 


\section*{Acknowledgments}
We thank Swiss National Science Foundation (SNSF) and Platform for Advanced Scientific Computing (PASC) for funding this project (project no. 205321\_173016), as well as Nvidia for GPU donations.





\bibliographystyle{ieee}

\bibliography{refs}

\end{document}